\documentclass[aps,prl,twocolumn,showpacs,amsmath,floatfix,superscriptaddress]{revtex4}

\usepackage{graphicx}
\usepackage{color}
\usepackage{amsmath}
\usepackage{dcolumn}
\usepackage{bm}
\usepackage{setspace}
\usepackage{mhchem}

\begin{document}

\title{Carrier Plasmon Induced Nonlinear Band Gap Renormalization in Two-Dimensional Semiconductors }

\author{Yufeng Liang, Li Yang}
\affiliation{Department of Physics, Washington University in St.
Louis, St. Louis, MO 63130, USA}

\date{\today}

\begin{abstract}
In reduced-dimensional semiconductors, doping-induced carrier
plasmons can strongly couple with quasiparticle excitations,
leading to a significant band gap renormalization. This effect has
been long known and is essential for understanding transport and
optical properties. However, the physical origin of this generic
effect remains obscure. We develop a new plasmon-pole theory that
efficiently and accurately capture this coupling. Using monolayer
\ce{MoS2} as a prototype two-dimensional (2D) semiconductor, we
reveal a striking band gap renormalization around 400 meV and an
unusual nonlinear evolution of its band gap with doping. This 2D
prediction significantly differs from the linear behaviors that
are common to one-dimensional structures. Our developed approach
allows for a quantitative understanding of many-body interactions
in general doped 2D semiconductors and paves the way for novel
band gap engineering techniques.
\end{abstract}

\maketitle

The band gap is a defining property of semiconductors and it is
typically not strongly influenced by extrinsic factors, such as
doping. However, because of dramatically enhanced many-electron
interactions, \cite{spataru2004excitonic,yang2007quasiparticle},
doped one-dimensional (1D) structures exhibit an unexpectedly
large band gap normalization (BGR) around several hundreds meV
\cite{spataru2010tunable,spataru2013quasiparticle}. Recently,
graphene-inspired two-dimensional (2D) semiconductors and their
excited-state properties have garnered enormous interest
\cite{mak2010atomically,splendiani2010emerging,mak2012control,wang2012electronics,chhowalla2013chemistry,xiao2012coupled}.
Since doping is a common occurrence
\cite{radisavljevic2011single,mak2013tightly,radisavljevic2013mobility,mouri2013tunable},
understanding the effects of BGR is essential for interpreting
experimental measurements, such as angle-resolved photoemission
spectroscopy (ARPES) \cite{jin2013direct, eknapakul2014electronic}
and extracting exciton and trion
\cite{mak2013tightly,mouri2013tunable} binding energies.


Beyond immediate practical applications, obtaining accurate
quasiparticle (QP) energies and the corresponding band gap in
doped reduced-dimensional semiconductors stands as a fundamental
challenge. A particular difficulty is capturing the screening that
are dominated by a unique low-energy acoustic carrier plasmon,
\cite{ando1982electronic,sarma1996dynamical}. Unlike in an undoped
semiconductor, here the carrier plasmon strongly couples with QP
excitations. This results in an enhanced BGR and can even lead to
a satellite structure in the QP spectral function
\cite{guzzo2011valence,lischner2013physical,lischner2014satellite}.
The QP-plasmon coupling also results in a nonlinear resonance
profile in the self-energy
\cite{spataru2013quasiparticle,lischner2014satellite} that
complicates the solutions to the Dyson equation. In short, this
subtle but important carrier plasmon calls for a special dynamical
treatment of the dielectric screening that is beyond the scope of
the widely used general plasmon pole (GPP) model
\cite{hybertsen1986electron} but is crucial for understanding
electronci structure of general 2D semiconductors.

In this work, we focus on a material of broad interest, monolayer
\ce{MoS2} and study its BGR over a wide range of doping densities
($n_\mathrm{2D}$) using the $GW$ approximation. We propose and
implement a generic plasmon-pole model (PPM) approach that
captures the essential screening effect and markedly improves the
efficiency of many-body calculations. Our study reveals that the
QP band gap of \ce{MoS2} exhibits a strongly nonlinear evolution
when varying the doping density; it drops sharply from 2.7 eV
across low doping densities but nearly saturates at 2.3 eV for
high densities. This is a consequence of the delicate interplay
between carrier occupation and dielectric screening. Ultimately,
beyond the one-shot $G_0W_0$ approximation
\cite{teinhoff2014influence}, we show that inclusion of
self-consistency is crucial for producing reliable QP energies in
the presence of strong QP-plasmon coupling.

The quasiparticle self-energy can be obtained using the GW
approximation \cite{hedin1965new}, \emph{i.e.}, $\Sigma=iGW$,
where $G$ is the single-particle propagator and $W$ is the
screened Coulomb interaction. For doped materials, the self-energy
can be decomposed into four terms \cite{spataru2013quasiparticle}
\begin{equation}
\begin{aligned}
\Sigma
&=i(G_\mathrm{int}W_\mathrm{int}
+\delta G W_\mathrm{int}
+G_\mathrm{int}\delta W
+\delta G\delta W)\\
&=i(\Sigma_\mathrm{int}+\Sigma_1+\Sigma_2+\Sigma_3)
\end{aligned}
\label{GW}
\end{equation}
where the subscript \lq\lq int\rq\rq denotes the operator of the
intrinsic (undoped) system and the $\delta$ terms capture the full
effects of the doping. The primary goal, then, is to find the
variation in dielectric screening $\delta\epsilon^{-1}$ and hence
$\delta W=\delta\epsilon^{-1}v$.

\emph{Dielectric function.} For a suspended 2D crystal structure,
the dielectric function $\varepsilon$ can be calculated using the
plane-wave representation \cite{hybertsen1986electron}:
$\epsilon_{\textbf{GG}'}(\textbf{q},\omega)=\delta_{\textbf{GG}'}-v_\mathrm{2D}(\textbf{q}+\textbf{G})\chi_{\textbf{GG}'}(\textbf{q},\omega)
\label{eps}$,
where $v_\mathrm{2D}(\textbf{q})$ is the 2D-truncated Coulomb
interaction \cite{rozzi2006exact,ismail2006truncation} and the
polarizability $\chi$ is obtained using the random phase
approximation. For a semiconductor with a sizeable band gap, the
dynamical matrix $\epsilon^{-1}_{\textbf{GG}'}(\textbf{q},\omega)$
is often described via the generalized plasmon-pole (GPP) model
with a simple-pole function that is based on the static limit and
$f$-sum rule \cite{hybertsen1986electron}.

A full-frequency calculation of the dielectric function may give
the accurate band gap of doped 2D semiconductors. However,
mimicking experimentally-accessible doping densities requires one
to use an ultra fine sampling scheme over the
$\textbf{k}$-$\omega$ space to capture the carrier screening,
making the simulation formidable for converged results. We are
thus motivated to develop an efficient and accurate \emph{ab
initio} model.

\begin{figure}
\centering
\includegraphics[width=0.5\textwidth]{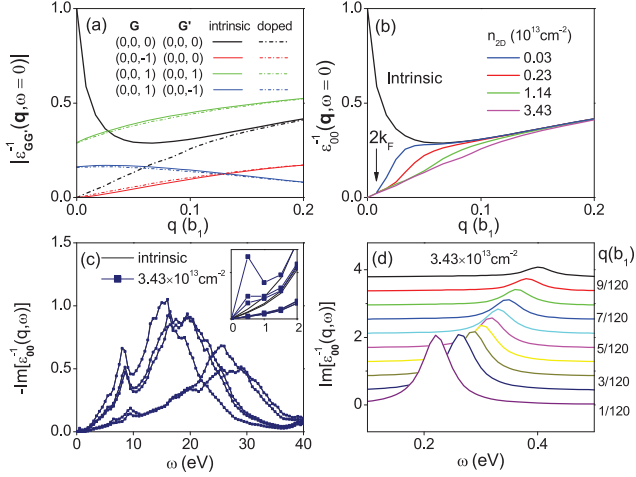}
\caption{(color online) (a) Static $\epsilon^{-1}_{\textbf{GG}'}$
with different combination of $\textbf{G}$-vectors. (b) Static
$\epsilon^{-1}_{\textbf{00}'}$ with at various doping density
$n_\mathrm{2D}$. (c) Loss function
$\mathrm{Im}[\epsilon^{-1}_{\textbf{00}'}(\textbf{q},\omega)]$ for
wave vector $\textbf{q}$ some high-symmetry points. (d) Loss
function at low energies featured by a shifting carrier-plasmon
peak.
} \label{Fig_1}
\end{figure}

Our model begins with the static
$\epsilon^{-1}_{\textbf{GG}'}(\textbf{q},\omega=0)$ of doped
monolayer \ce{MoS2}. Fig. \ref{Fig_1} (a) compares several
representative dielectric matrix elements of the undoped system
and the doped one at $n_\mathrm{2D} =3.4\times10^{13}$cm$^{-2}$.
Given the isotropy about small $q$ \cite{supp}, the elements are
only plotted along a single reciprocal primitive vector
$\textbf{b}_1$. Remarkably, all of the matrix elements are remain
nearly unaffected (difference $<0.01$) even at this high doping
level, except the \lq\lq head\rq\rq matrix element
$\epsilon^{-1}_{\textbf{00}}(\textbf{q},0)$. In the undoped case,
$\epsilon^{-1}_{\mathrm{int},\textbf{00}}(\textbf{q},0)$
approaches $1$ \cite{ismail2006truncation,deslippe2012berkeleygw}
as $q\rightarrow 0$, reflecting the absence of screening in a 2D
semiconductor at long-wavelengths. However, including doping
causes $\epsilon^{-1}_{\textbf{00}}(\textbf{q},0)$ immediately
drops to $0$ \cite{deslippe2012berkeleygw}, due to metallic
screening. This result leads us to focus primarily on the
variation of the head matrix element
$\delta\epsilon^{-1}_{\textbf{00}}$.

$\epsilon^{-1}_{\textbf{00}}(\textbf{q},0)$ is shown in Fig.
\ref{Fig_1} (b) at various doping densities $n_\mathrm{2D}$. For
increasing q, all of the dielectric functions
$\epsilon^{-1}_{\textbf{00}}(\textbf{q},0)$ at finite
$n_\mathrm{2D}$ first grow up linearly at the same rate for small
q, but then individually turn up towards, and merge with,  the
intrinsic
$\epsilon^{-1}_{\mathrm{int},\textbf{00}}(\textbf{q},0)$.
Interestingly, the turning point for each density is located at
$q=2k_F$, where $k_F=\sqrt{2\pi n_\mathrm{2D}}$ is the Fermi wave
Vector for a given doping density. This behavior is in accord with
the 2D free electron gas (FEG)
polarizability\cite{giuliani2005quantum},
$\delta\chi_{\textbf{00}}(\textbf{q})=-\frac{m^*}{2\pi}[1-\theta(q-2k_F)\sqrt{1-4k^2_F/q^2}]$,
for which the static doping effect only dominates at small $q$ and
damps away rapidly beyond $2k_F$.

Next, we turn to dynamical screening effects to treat the carrier
plasmon. Fig. \ref{Fig_1} (c) presents the frequency-dependence of
the loss function
$\mathrm{Im}[\epsilon^{-1}_{\textbf{00}}(\textbf{q},\omega)]$ for
various $\textbf{q}$ located at several high-symmetry points of
the first Brillouin zone (BZ). It can be seen that the intrinsic
and doped cases are almost identical ($<0.1\%$) except for some
spectral features at low frequencies (inset of Fig.
\ref{Fig_1}(c)). A refined calculation (Fig. \ref{Fig_1}(d))
provides clear evidence of a dispersive carrier plasmon, as
manifested by the peak in the loss function.

\begin{figure}
\centering
\includegraphics[width=0.5\textwidth]{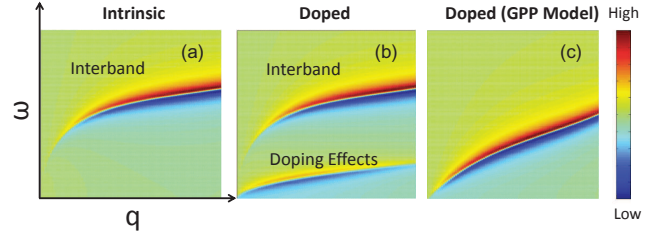}
\caption{(color online). Schematics of
$\mathrm{Re}[\epsilon^{-1}_\textbf{00}](\textbf{q},\omega)$
produced by simple-pole functions, as indicated by colormap plot
on the $q-\omega$ plane. (a) Intrinsic case, (b) doped case, and
(c) GPP model for doped case.} \label{Fig_2}
\end{figure}

The above calculation reveals a key fact about the dielectric
function of a doped 2D system: \emph{the doping effects are
exclusively concentrated in the head element
$\epsilon^{-1}_{\textbf{00}}(q,\omega)$ at long wavelengths and
low frequencies; this is where the carrier plasmon dominates}. An
overall picture is illustrated with schematics in Fig.
\ref{Fig_2}. For an undoped semiconductor, only the optical
plasmons arising from interband transitions are present, which can
be represented by the single-pole function (Fig. \ref{Fig_2}(a)).
For a doped 2D semiconductor, a branch of low-energy carrier
acoustic plasmon emerges  while the high-energy optical plasmon
remains intact (Fig. 2(b)). The GPP model no longer accurately
describes the dynamical effects in the system when this new
plasmon emerges; it exaggerates the doping effects across a
broader frequency region (Fig. \ref{Fig_2}(c)).

Motivated by the simple plasmon structure in Fig. \ref{Fig_1}(d),
we model the variation of the head matrix element caused by doping
as
\begin{equation}
\delta\epsilon^{-1}_{\textbf{00}}(\textbf{q}, \omega)
=\frac{\Omega^2_d(\textbf{q})}{\omega^2-\omega^2_d(\textbf{q})}
\label{PPM}
\end{equation}
where the parameter $\Omega_d(\textbf{q})$ and
$\omega_d(\textbf{q})$ are the plasmon-pole strength and
frequency, respectively. These are determined by the following two
constraints. First, the plasmon energy $\omega_d(\textbf{q})$ can
be extracted from a frequency-dependent calculation. We find that
$\omega_d(\textbf{q})$ converges quickly with cutoff energy
$E_\textbf{G}$ and occupied band number $N_c$. In our case,
$E_\textbf{G}=2$ Ry and $N_c=4$ are sufficient for producing a
converged $\omega_d(\textbf{q})$ \cite{supp}. Second, the
plasmon-pole strength is given by
$\Omega^2_d(\textbf{q})=-\delta\epsilon^{-1}_\textbf{00}(\textbf{q},\omega=0)\omega^2_d(\textbf{q})$,
where $\delta\epsilon^{-1}$ can be extracted from static
calculations for the intrinsic and doped systems (Fig. 1(a)).
Although a dense $\textbf{k}$-grid ($120\times120\times1$) and a
typical  $E_\textbf{G}$ ($10 $Ry) are needed, only a few
conduction bands ($N_c=13$) are needed for convergence for
calculations involving only small $q$.

\begin{figure}
\centering
\includegraphics[width=0.5\textwidth]{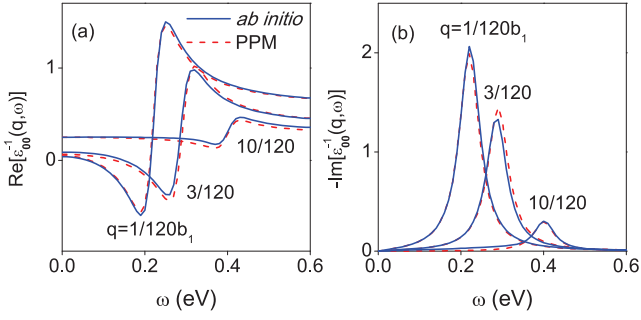}
\caption{(color online) Real and imaginary parts of the variation
in the head matrix element, $\delta\epsilon^{-1}_\textbf{00}$, as
obtained from full \emph{ab initio} calculations and our proposed
PPM. } \label{Fig_3}
\end{figure}

As shown in Fig. \ref{Fig_3}, this PPM satisfactorily reproduces
the frequency-dependence of the head $\epsilon^{-1}_{\textbf{00}}$
that are obtained from \emph{ab initio} simulations, for various
wave vector $\textbf{q}$. Our proposed calculation scheme focuses
on the head dielectric function at small wave vectors and low
frequencies, which circumvents the process of inverting the
dielectric matrix \cite{oschlies1995gw}. Ultimately, this model is
a far more efficient than the full-frequency scheme.

\emph{Self-Energy}: We will calculate the self-energy of doped 2D
semiconductors using the developed PPM. Following the COHSEX
approximation \cite{hedin1965new,hybertsen1986electron}, the
self-energy $\Sigma^{n\textbf{k}}$ can be split into a
screened-exchange (SX) term and a Coulomb-hole (CH) term
\begin{equation}
\begin{aligned}
\Sigma^{n\textbf{k}}_{\mathrm{SX/CH}}(E)=
\sum\limits_{n'\textbf{q},\textbf{GG}'}
\xi^{n'n}_{-\textbf{G},-\textbf{G}'}(\textbf{k},-\textbf{q})K_{\mathrm{SX/CH}}
\end{aligned}
\label{COHSEX}
\end{equation}
where the SX and CH kernels are defined as $K_{\mathrm{SX}}=
-f_{n'\textbf{k}-\textbf{q}}
W_{\textbf{GG}'}(\textbf{q},E-\varepsilon_{n'\textbf{k}-\textbf{q}})$
and $K_{\mathrm{CH}} =
W^{+}_{\textbf{GG}'}(\textbf{q},E-\varepsilon_{n'\textbf{k}-\textbf{q}})$,
with
$W^\pm(E)=\pm\frac{1}{\pi}\mathrm{P}\int\limits^{\pm\infty}_{0}\mathrm{d}E'\frac{\mathrm{Im}W(E)}{E-E'}$,
which only encompasses the positive (negative) poles in $W$.
$\xi^{nn'}_{\textbf{GG}'}(\textbf{k},\textbf{q})
=M^*_{nn'}(\textbf{k},\textbf{q},\textbf{G})
M_{nn'}(\textbf{k},\textbf{q},\textbf{G}')$ represents the band
structure effect, where
$M_{nn'}(\textbf{k},\textbf{q},\textbf{G})$ is the plane-wave
matrix element\cite{hybertsen1986electron,deslippe2012berkeleygw}.
$\varepsilon_{n\textbf{k}}$ and $f_{n\textbf{k}}$ are the
single-particle energy and occupation number of the state on band
$n$ at $\textbf{k}$, respectively.

According to Eq. (\ref{GW}), we can obtain the total self-energy
term by term. The calculation of $\Sigma_\mathrm{int}$ and
$\Sigma_1$ is straightforward
\cite{qiu2013optical,liang2013quasiparticle,supp}. The next two
terms, $\Sigma_2$ and $\Sigma_3$, are closely tied to the carrier
screening $\delta W$, which can be reproduced using our proposed
PPM in Eq. (\ref{PPM}). Due to the aforementioned properties of
the carrier plasmon, these self-energy contributions can be
simplified dramatically by (i) limiting the band summation to
$n'=n$, (ii) retaining only the term with
$\textbf{G}=\textbf{G}'=\textbf{0}$, and (iii) setting up a cutoff
$q_c$ for the BZ integration.

$\Sigma_2=iG_\mathrm{int}\delta W$ does not involve partial band
filling, it reads
\begin{equation}
\begin{aligned}
\Sigma^{n\textbf{k}}_{2}(E)\simeq&
\pm\int\limits_{q<q_c}
\frac{\mathrm{d}^2\textbf{q}}{(2\pi)^2}
\xi^{nn}_{\textbf{00}}(\textbf{k},-\textbf{q})
\delta W^\pm_{\textbf{00}}(\textbf{q},E-\varepsilon_{n\textbf{k}-\textbf{q}})
\end{aligned}
\label{v}
\end{equation}
where $\delta W^\pm_{\textbf{00}}(\textbf{q},\omega)
=\pm\dfrac{\Omega^2_d(\textbf{q})}
{2\omega_d(\textbf{q})(\omega\mp\omega_d(\textbf{q}))}v_\mathrm{2D}(\textbf{q})
\label{dW}$ and $\pm$ is for conduction/valence states. For fully
filled valence bands, the SX and CH term have been combined via
$-W^-=-W+W^+$. $\Sigma_3=i\delta G\delta W$ is affected by the
carrier occupation and is solely related to the SX term
\begin{equation}
\begin{aligned}
\Sigma^{n\textbf{k}}_{3}(E)\simeq
-\int\limits_{q<q_c}
\frac{\mathrm{d}^2\textbf{q}}{(2\pi)^2}
&\xi^{nn}_{\textbf{00}}(\textbf{k},-\textbf{q})\\
\times
&\delta f_{n\textbf{k}-\textbf{q}}
\delta W_{\textbf{00}}(\textbf{q},E-\varepsilon_{n\textbf{k}-\textbf{q}})
\end{aligned}
\end{equation}
This term is only significant on the doped band.

\begin{figure}
\centering
\includegraphics[width=0.5\textwidth]{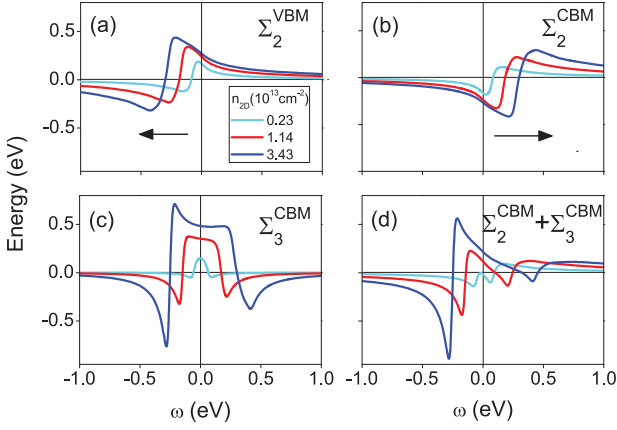}
\caption{(color online) Self-energy contributions
$\Sigma_2=iG_\mathrm{int}\delta W$ and $\Sigma_3=i\delta G\delta
W$ for both VBM and CBM at different doping levels. $\omega=0$ is
set at the on-shell energy. The arrows in (a) and (b) are used for
guiding readers' eyes for the evolution of those resonant peaks
under higher doping densities.} \label{Fig_4}
\end{figure}

Finally, the energy dependence of the self-energy plays a crucial
role in determining the QP energies. For a state ($n\textbf{k}$),
both $\Sigma_\mathrm{int}$ and $\Sigma_1$ vary slowly near the
single-particle energy $\varepsilon_{n\textbf{k}}$ because the
optical-plasmon feature occurs at high-energies \cite{supp}.
However, this is not the case for $\Sigma_2$ and $\Sigma_3$, on
account of the emergence of the low-energy carrier plasmon.
Instead, they exhibit strongly nonlinear behaviors near the
$\varepsilon_{n\textbf{k}}$. Fig. \ref{Fig_4} displays $\Sigma_2$
and $\Sigma_3$ for the valence band maximum (VBM) and conduction
band minimum (CBM) of n-doped monolayer \ce{MoS2} with
$\omega=E-\varepsilon_\mathrm{VBM}$ and
$\omega=E-\varepsilon_\mathrm{CBM}$, respectively. Thus $\omega=0$
defines the on-shell energy. For simplicity, spin-orbital coupling
is not considered here. In Fig. \ref{Fig_4} (a),
$\Sigma^{\mathrm{VBM}}_2$ exhibits typical feature of Fano
resonance, resulting from the coupling of a quasi-electron with
the dispersive carrier plasmon. With increasing carrier density
$n_\mathrm{2D}$, the resonance peak position systematically shifts
left due to a blue shift in the plasmon energy. The case of
$\Sigma^{\mathrm{CBM}}_2$ is the reverse (Fig. \ref{Fig_4} (b))
because it corresponds to a quasi-hole state. Fig. \ref{Fig_4} (c)
shows the energy dependence of $\Sigma_3$. For electron doping,
while $\Sigma^{\mathrm{VBM}}_3\simeq0$, $\Sigma^{\mathrm{CBM}}_3$
rises as an asymmetric finite-width plateau. Finally, the total
contribution of $\Sigma^{\mathrm{CBM}}_2+\Sigma^{\mathrm{CBM}}_3$
is shown in Fig. \ref{Fig_4}(d), which features a transition from
a hole-like resonance to an electron-like resonance.

With all of the self-energy contributions taken under
consideration, the QP band energies $E$ can be obtained by solving
the Dyson equation \cite{hybertsen1986electron}
$E_{n\textbf{k}}=\varepsilon_{n\textbf{k}}-V^{n\textbf{k}}_\mathrm{XC}+\mathrm{Re}[\Sigma^{n\textbf{k}}(E_{n\textbf{k}})]$
where Vxc is the exchange-correlation potential from the DFT.
Although the $G_0W_0$ approximation works well for intrinsic
semiconductors, it is no longer appropriate for treating a doped
2D system, for which the self-energy becomes strongly nonlinear.
This is manifested by $\Sigma_2$ and $\Sigma_3$. This nonlinear
profile causes the QP solutions to depend sensitively on the
preliminary DFT calculation. Even worse, now there can be multiple
solutions. One has to determine which is the most important one by
considering the spectral weight, which also relies on the DFT
starting-point. This has exacerbates the inherent deficiency of
the $G_0W_0$ approximation. One needs to proceed with a
self-consistent $GW$ calculation
\cite{bruneval2006effect,van2006quasiparticle}. Fortunately, we
find that the carrier-plasmon resonance profile depends weakly on
the band curvature \cite{supp}. This suggests that the
self-consistency can be achieved by rigidly shifting
\cite{hedin1999correlation} the whole resonance profile along
energy-axis such that the on-shell energy of $\Sigma$ coincides
with the QP solution. This procedure is equivalent to performing
the $GW_0$ approximation
\cite{lischner2013physical,sarma2013velocity}.

\begin{figure}
\centering
\includegraphics[width=0.5\textwidth]{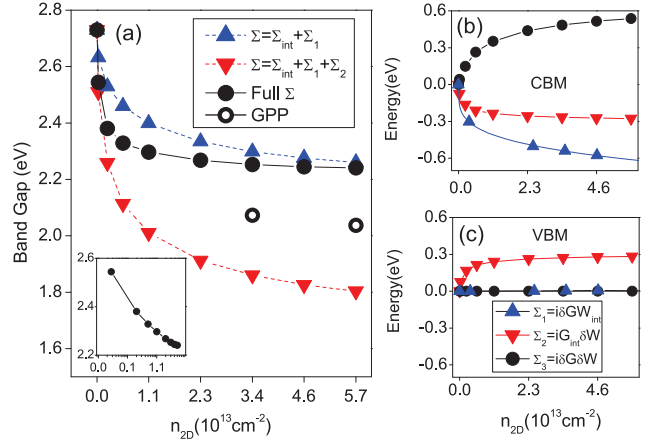}
\caption{(color online) (a) QP band gap evolution versus the
doping density. The impact of partial summation of self-energy
contribution on the band gap is also discussed. The GPP band gap
is marked by open circles. The inset is shown in logarithm scale
of $n_\mathrm{2D}$. (b)(c) on-shell self-energy corrections in CBM
(b) and VBM (c) from doping effects.} \label{Fig_5}
\end{figure}

Fig. \ref{Fig_5} displays the evolution of the BGR and the
different contributing mechanisms versus doping density
$n_\mathrm{2D}$ in monolayer \ce{MoS2}. Remarkably, the QP band
gap is strongly renormalized in the light-doping regime; it drops
dramatically from $2.73$ eV to around $2.25$ eV as $n_\mathrm{2D}$
increases from 0 to $10^{13}$cm$^{-2}$ (Fig. \ref{Fig_5}(a)). The
origin of this enhanced band-gap shrinkage is due to the following
mechanisms. First, a major contribution comes from the
carrier-occupation energy. At low enough doping density, this
contribution is dominated by $\Sigma_1=i\delta G\cdot
W_\mathrm{int}$, which is analogous to the \emph{negative}
Hartree-Fock exchange energy\cite{giuliani2005quantum}, except
that the bare Coulomb interaction $v$ is replaced by
$W_\mathrm{int}$. Thus for a n-doped system of any dimensionality,
$\Sigma_1$ roughly scales as $-\epsilon_\mathrm{int}^{-1} k_F$,
which always lowers the CBM (Fig. \ref{Fig_5} (b)). Since the
screening effect $\epsilon_\mathrm{int}$ is greatly depressed in
low-dimensional semiconductors
\cite{yang2007quasiparticle,yang2009excitonic,spataru2004excitonic},
the band gap reduction from the carrier occupation, $\Sigma_1$, is
particularly large. Within the largest $n_\mathrm{2D}$ in our
simulation, it can reach up to $\sim500$meV. Similar magnitudes of
carrier-occupation energy have also been reported in doped
semiconducting carbon nanotubes (CNTs)
\cite{spataru2010tunable,spataru2013quasiparticle}. Another
important band-gap shrinkage mechanism comes from the carrier
screening $\Sigma_2=iG_\mathrm{int}\delta W$. It weakens the
electron-electron interaction and hence leads to a significant
reduction in QP band gap correction up to a few hundred meV (see
Fig. \ref{Fig_5} (b) and (c)).

As the doping density $n_\mathrm{2D}$ increases, the QP band gap
saturates at $2.25$eV eventually. This results from several
factors. When $n_\mathrm{2D}$ is high enough, the extra carrier
screening can in turn reduce the carrier occupation energy. At
high $n_\mathrm{2D}$, the double-difference term $\Sigma_3=i\delta
G\delta W$ becomes a prominent contribution (Fig. \ref{Fig_5}
(b)). For n-doping, it scales as $-\delta\epsilon^{-1}k_F$.
Therefore, $\Sigma_3$ varies in an opposite trend to $\Sigma_1$
due to this minus sign and is responsible for enlarging the band
gap. Furthermore, the carrier screening effect $\Sigma_2$ itself
also exhibits a saturation behavior (Fig. \ref{Fig_5} (b) and
(c)). With increasing doping density, the carrier plasmon
blueshifts significantly and its effect on the QP state becomes
weaker, as evidenced from the departing Fano feature that is
indicated by arrows in Fig \ref{Fig_4} (a) and (b). Finally,
dimensionality effects also plays a role in the band gap
evolution. Our case differs substantially from that in CNTs, for
which the band gap shrinks almost linearly with the doping density
\cite{spataru2010tunable}. In the 1D case, some major self-energy
contributions, like $\Sigma_1$, still scale as $-k_F$, but which
is proportional to the doping density.

For comparison, we calculate the band gap via the conventional GPP
method for $n_\mathrm{2D}=3.4\times10^{13}$ and
$5.7\times10^{13}$cm$^{-2}$. As shown in Fig. \ref{Fig_5} (a), the
GPP model overestimates the band gap reduction by $\simeq200$meV,
which is $40\%$ larger than in our refined PPM method.

In conclusion, we have developed and applied a highly efficient
generic computing scheme for calculating the dielectric function
of doped 2D semiconductors. From it, we obtained the enhanced and
nonlinear reduction of the QP band gap over a wide range of doping
densities. This has not been observed in one-dimensional and bulk
structures. We also found that the band gap drops to a certain
limit at sufficiently high doping densities because of a delicate
competition between the exchange and correlation energies. This
enhanced BGR is crucial for explaining the excitonic effects and
trions observed in experiments and it can be directly observed by
a combination of the APRES and inverse ARPES spectroscopies.

We acknowledge valuable discussions with Giovanni Vignale and Ryan
Soklaski. This work is supported by NSF DMR-1207141. The
ground-state calculation is performed with Quantum Espresso
\cite{giannozzi2009quantum}. The intrinsic self-energies are
obtained by the BerkeleyGW code \cite{deslippe2012berkeleygw}. The
computational resources have been provided by Lonestar of Teragrid
at the Texas Advanced Computing Center (TACC).


\begin{thebibliography}{38}
\expandafter\ifx\csname natexlab\endcsname\relax\def\natexlab#1{#1}\fi
\expandafter\ifx\csname bibnamefont\endcsname\relax
  \def\bibnamefont#1{#1}\fi
\expandafter\ifx\csname bibfnamefont\endcsname\relax
  \def\bibfnamefont#1{#1}\fi
\expandafter\ifx\csname citenamefont\endcsname\relax
  \def\citenamefont#1{#1}\fi
\expandafter\ifx\csname url\endcsname\relax
  \def\url#1{\texttt{#1}}\fi
\expandafter\ifx\csname urlprefix\endcsname\relax\def\urlprefix{URL }\fi
\providecommand{\bibinfo}[2]{#2}
\providecommand{\eprint}[2][]{\url{#2}}

\bibitem[{\citenamefont{Spataru et~al.}(2004)\citenamefont{Spataru,
  Ismail-Beigi, Benedict, and Louie}}]{spataru2004excitonic}
\bibinfo{author}{\bibfnamefont{C.~D.} \bibnamefont{Spataru}},
  \bibinfo{author}{\bibfnamefont{S.}~\bibnamefont{Ismail-Beigi}},
  \bibinfo{author}{\bibfnamefont{L.~X.} \bibnamefont{Benedict}},
  \bibnamefont{and} \bibinfo{author}{\bibfnamefont{S.~G.} \bibnamefont{Louie}},
  \bibinfo{journal}{Physical Review Letters} \textbf{\bibinfo{volume}{92}},
  \bibinfo{pages}{077402} (\bibinfo{year}{2004}).

\bibitem[{\citenamefont{Yang et~al.}(2007)\citenamefont{Yang, Park, Son, Cohen,
  and Louie}}]{yang2007quasiparticle}
\bibinfo{author}{\bibfnamefont{L.}~\bibnamefont{Yang}},
  \bibinfo{author}{\bibfnamefont{C.-H.} \bibnamefont{Park}},
  \bibinfo{author}{\bibfnamefont{Y.-W.} \bibnamefont{Son}},
  \bibinfo{author}{\bibfnamefont{M.~L.} \bibnamefont{Cohen}}, \bibnamefont{and}
  \bibinfo{author}{\bibfnamefont{S.~G.} \bibnamefont{Louie}},
  \bibinfo{journal}{Physical Review Letters} \textbf{\bibinfo{volume}{99}},
  \bibinfo{pages}{186801} (\bibinfo{year}{2007}).

\bibitem[{\citenamefont{Spataru and L{\'e}onard}(2010)}]{spataru2010tunable}
\bibinfo{author}{\bibfnamefont{C.~D.} \bibnamefont{Spataru}} \bibnamefont{and}
  \bibinfo{author}{\bibfnamefont{F.}~\bibnamefont{L{\'e}onard}},
  \bibinfo{journal}{Physical review letters} \textbf{\bibinfo{volume}{104}},
  \bibinfo{pages}{177402} (\bibinfo{year}{2010}).

\bibitem[{\citenamefont{Spataru and
  L{\'e}onard}(2013)}]{spataru2013quasiparticle}
\bibinfo{author}{\bibfnamefont{C.~D.} \bibnamefont{Spataru}} \bibnamefont{and}
  \bibinfo{author}{\bibfnamefont{F.}~\bibnamefont{L{\'e}onard}},
  \bibinfo{journal}{Chemical Physics} \textbf{\bibinfo{volume}{413}},
  \bibinfo{pages}{81} (\bibinfo{year}{2013}).

\bibitem[{\citenamefont{Mak et~al.}(2010)\citenamefont{Mak, Lee, Hone, Shan,
  and Heinz}}]{mak2010atomically}
\bibinfo{author}{\bibfnamefont{K.~F.} \bibnamefont{Mak}},
  \bibinfo{author}{\bibfnamefont{C.}~\bibnamefont{Lee}},
  \bibinfo{author}{\bibfnamefont{J.}~\bibnamefont{Hone}},
  \bibinfo{author}{\bibfnamefont{J.}~\bibnamefont{Shan}}, \bibnamefont{and}
  \bibinfo{author}{\bibfnamefont{T.~F.} \bibnamefont{Heinz}},
  \bibinfo{journal}{Physical Review Letters} \textbf{\bibinfo{volume}{105}},
  \bibinfo{pages}{136805} (\bibinfo{year}{2010}).

\bibitem[{\citenamefont{Splendiani et~al.}(2010)\citenamefont{Splendiani, Sun,
  Zhang, Li, Kim, Chim, Galli, and Wang}}]{splendiani2010emerging}
\bibinfo{author}{\bibfnamefont{A.}~\bibnamefont{Splendiani}},
  \bibinfo{author}{\bibfnamefont{L.}~\bibnamefont{Sun}},
  \bibinfo{author}{\bibfnamefont{Y.}~\bibnamefont{Zhang}},
  \bibinfo{author}{\bibfnamefont{T.}~\bibnamefont{Li}},
  \bibinfo{author}{\bibfnamefont{J.}~\bibnamefont{Kim}},
  \bibinfo{author}{\bibfnamefont{C.-Y.} \bibnamefont{Chim}},
  \bibinfo{author}{\bibfnamefont{G.}~\bibnamefont{Galli}}, \bibnamefont{and}
  \bibinfo{author}{\bibfnamefont{F.}~\bibnamefont{Wang}},
  \bibinfo{journal}{Nano letters} \textbf{\bibinfo{volume}{10}},
  \bibinfo{pages}{1271} (\bibinfo{year}{2010}).

\bibitem[{\citenamefont{Mak et~al.}(2012)\citenamefont{Mak, He, Shan, and
  Heinz}}]{mak2012control}
\bibinfo{author}{\bibfnamefont{K.~F.} \bibnamefont{Mak}},
  \bibinfo{author}{\bibfnamefont{K.}~\bibnamefont{He}},
  \bibinfo{author}{\bibfnamefont{J.}~\bibnamefont{Shan}}, \bibnamefont{and}
  \bibinfo{author}{\bibfnamefont{T.~F.} \bibnamefont{Heinz}},
  \bibinfo{journal}{Nature nanotechnology} \textbf{\bibinfo{volume}{7}},
  \bibinfo{pages}{494} (\bibinfo{year}{2012}).

\bibitem[{\citenamefont{Wang et~al.}(2012)\citenamefont{Wang, Kalantar-Zadeh,
  Kis, Coleman, and Strano}}]{wang2012electronics}
\bibinfo{author}{\bibfnamefont{Q.~H.} \bibnamefont{Wang}},
  \bibinfo{author}{\bibfnamefont{K.}~\bibnamefont{Kalantar-Zadeh}},
  \bibinfo{author}{\bibfnamefont{A.}~\bibnamefont{Kis}},
  \bibinfo{author}{\bibfnamefont{J.~N.} \bibnamefont{Coleman}},
  \bibnamefont{and} \bibinfo{author}{\bibfnamefont{M.~S.}
  \bibnamefont{Strano}}, \bibinfo{journal}{Nature nanotechnology}
  \textbf{\bibinfo{volume}{7}}, \bibinfo{pages}{699} (\bibinfo{year}{2012}).

\bibitem[{\citenamefont{Chhowalla et~al.}(2013)\citenamefont{Chhowalla, Shin,
  Eda, Li, Loh, and Zhang}}]{chhowalla2013chemistry}
\bibinfo{author}{\bibfnamefont{M.}~\bibnamefont{Chhowalla}},
  \bibinfo{author}{\bibfnamefont{H.~S.} \bibnamefont{Shin}},
  \bibinfo{author}{\bibfnamefont{G.}~\bibnamefont{Eda}},
  \bibinfo{author}{\bibfnamefont{L.-J.} \bibnamefont{Li}},
  \bibinfo{author}{\bibfnamefont{K.~P.} \bibnamefont{Loh}}, \bibnamefont{and}
  \bibinfo{author}{\bibfnamefont{H.}~\bibnamefont{Zhang}},
  \bibinfo{journal}{Nature chemistry} \textbf{\bibinfo{volume}{5}},
  \bibinfo{pages}{263} (\bibinfo{year}{2013}).

\bibitem[{\citenamefont{Xiao et~al.}(2012)\citenamefont{Xiao, Liu, Feng, Xu,
  and Yao}}]{xiao2012coupled}
\bibinfo{author}{\bibfnamefont{D.}~\bibnamefont{Xiao}},
  \bibinfo{author}{\bibfnamefont{G.-B.} \bibnamefont{Liu}},
  \bibinfo{author}{\bibfnamefont{W.}~\bibnamefont{Feng}},
  \bibinfo{author}{\bibfnamefont{X.}~\bibnamefont{Xu}}, \bibnamefont{and}
  \bibinfo{author}{\bibfnamefont{W.}~\bibnamefont{Yao}},
  \bibinfo{journal}{Physical Review Letters} \textbf{\bibinfo{volume}{108}},
  \bibinfo{pages}{196802} (\bibinfo{year}{2012}).

\bibitem[{\citenamefont{Radisavljevic et~al.}(2011)\citenamefont{Radisavljevic,
  Radenovic, Brivio, Giacometti, and Kis}}]{radisavljevic2011single}
\bibinfo{author}{\bibfnamefont{B.}~\bibnamefont{Radisavljevic}},
  \bibinfo{author}{\bibfnamefont{A.}~\bibnamefont{Radenovic}},
  \bibinfo{author}{\bibfnamefont{J.}~\bibnamefont{Brivio}},
  \bibinfo{author}{\bibfnamefont{V.}~\bibnamefont{Giacometti}},
  \bibnamefont{and} \bibinfo{author}{\bibfnamefont{A.}~\bibnamefont{Kis}},
  \bibinfo{journal}{Nature nanotechnology} \textbf{\bibinfo{volume}{6}},
  \bibinfo{pages}{147} (\bibinfo{year}{2011}).

\bibitem[{\citenamefont{Mak et~al.}(2013)\citenamefont{Mak, He, Lee, Lee, Hone,
  Heinz, and Shan}}]{mak2013tightly}
\bibinfo{author}{\bibfnamefont{K.~F.} \bibnamefont{Mak}},
  \bibinfo{author}{\bibfnamefont{K.}~\bibnamefont{He}},
  \bibinfo{author}{\bibfnamefont{C.}~\bibnamefont{Lee}},
  \bibinfo{author}{\bibfnamefont{G.~H.} \bibnamefont{Lee}},
  \bibinfo{author}{\bibfnamefont{J.}~\bibnamefont{Hone}},
  \bibinfo{author}{\bibfnamefont{T.~F.} \bibnamefont{Heinz}}, \bibnamefont{and}
  \bibinfo{author}{\bibfnamefont{J.}~\bibnamefont{Shan}},
  \bibinfo{journal}{Nature materials} \textbf{\bibinfo{volume}{12}},
  \bibinfo{pages}{207} (\bibinfo{year}{2013}).

\bibitem[{\citenamefont{Radisavljevic and
  Kis}(2013)}]{radisavljevic2013mobility}
\bibinfo{author}{\bibfnamefont{B.}~\bibnamefont{Radisavljevic}}
  \bibnamefont{and} \bibinfo{author}{\bibfnamefont{A.}~\bibnamefont{Kis}},
  \bibinfo{journal}{Nature materials} \textbf{\bibinfo{volume}{12}},
  \bibinfo{pages}{815} (\bibinfo{year}{2013}).

\bibitem[{\citenamefont{Mouri et~al.}(2013)\citenamefont{Mouri, Miyauchi, and
  Matsuda}}]{mouri2013tunable}
\bibinfo{author}{\bibfnamefont{S.}~\bibnamefont{Mouri}},
  \bibinfo{author}{\bibfnamefont{Y.}~\bibnamefont{Miyauchi}}, \bibnamefont{and}
  \bibinfo{author}{\bibfnamefont{K.}~\bibnamefont{Matsuda}},
  \bibinfo{journal}{Nano letters} \textbf{\bibinfo{volume}{13}},
  \bibinfo{pages}{5944} (\bibinfo{year}{2013}).

\bibitem[{\citenamefont{Jin et~al.}(2013)\citenamefont{Jin, Yeh, Zaki, Zhang,
  Sadowski, Al-Mahboob, van~der Zande, Chenet, Dadap, Herman
  et~al.}}]{jin2013direct}
\bibinfo{author}{\bibfnamefont{W.}~\bibnamefont{Jin}},
  \bibinfo{author}{\bibfnamefont{P.-C.} \bibnamefont{Yeh}},
  \bibinfo{author}{\bibfnamefont{N.}~\bibnamefont{Zaki}},
  \bibinfo{author}{\bibfnamefont{D.}~\bibnamefont{Zhang}},
  \bibinfo{author}{\bibfnamefont{J.~T.} \bibnamefont{Sadowski}},
  \bibinfo{author}{\bibfnamefont{A.}~\bibnamefont{Al-Mahboob}},
  \bibinfo{author}{\bibfnamefont{A.~M.} \bibnamefont{van~der Zande}},
  \bibinfo{author}{\bibfnamefont{D.~A.} \bibnamefont{Chenet}},
  \bibinfo{author}{\bibfnamefont{J.~I.} \bibnamefont{Dadap}},
  \bibinfo{author}{\bibfnamefont{I.~P.} \bibnamefont{Herman}},
  \bibnamefont{et~al.}, \bibinfo{journal}{Physical Review Letters}
  \textbf{\bibinfo{volume}{111}}, \bibinfo{pages}{106801}
  (\bibinfo{year}{2013}).

\bibitem[{\citenamefont{Eknapakul et~al.}(2014)\citenamefont{Eknapakul, King,
  Asakawa, Buaphet, He, Mo, Takagi, Shen, Baumberger, Sasagawa
  et~al.}}]{eknapakul2014electronic}
\bibinfo{author}{\bibfnamefont{T.}~\bibnamefont{Eknapakul}},
  \bibinfo{author}{\bibfnamefont{P.~D.} \bibnamefont{King}},
  \bibinfo{author}{\bibfnamefont{M.}~\bibnamefont{Asakawa}},
  \bibinfo{author}{\bibfnamefont{P.}~\bibnamefont{Buaphet}},
  \bibinfo{author}{\bibfnamefont{R.-H.} \bibnamefont{He}},
  \bibinfo{author}{\bibfnamefont{S.-K.} \bibnamefont{Mo}},
  \bibinfo{author}{\bibfnamefont{H.}~\bibnamefont{Takagi}},
  \bibinfo{author}{\bibfnamefont{K.~M.} \bibnamefont{Shen}},
  \bibinfo{author}{\bibfnamefont{F.}~\bibnamefont{Baumberger}},
  \bibinfo{author}{\bibfnamefont{T.}~\bibnamefont{Sasagawa}},
  \bibnamefont{et~al.}, \bibinfo{journal}{Nano letters}
  \textbf{\bibinfo{volume}{14}}, \bibinfo{pages}{1312} (\bibinfo{year}{2014}).

\bibitem[{\citenamefont{Ando et~al.}(1982)\citenamefont{Ando, Fowler, and
  Stern}}]{ando1982electronic}
\bibinfo{author}{\bibfnamefont{T.}~\bibnamefont{Ando}},
  \bibinfo{author}{\bibfnamefont{A.~B.} \bibnamefont{Fowler}},
  \bibnamefont{and} \bibinfo{author}{\bibfnamefont{F.}~\bibnamefont{Stern}},
  \bibinfo{journal}{Reviews of Modern Physics} \textbf{\bibinfo{volume}{54}},
  \bibinfo{pages}{437} (\bibinfo{year}{1982}).

\bibitem[{\citenamefont{Sarma and Hwang}(1996)}]{sarma1996dynamical}
\bibinfo{author}{\bibfnamefont{S.~D.} \bibnamefont{Sarma}} \bibnamefont{and}
  \bibinfo{author}{\bibfnamefont{E.}~\bibnamefont{Hwang}},
  \bibinfo{journal}{Physical Review B} \textbf{\bibinfo{volume}{54}},
  \bibinfo{pages}{1936} (\bibinfo{year}{1996}).

\bibitem[{\citenamefont{Guzzo et~al.}(2011)\citenamefont{Guzzo, Lani, Sottile,
  Romaniello, Gatti, Kas, Rehr, Silly, Sirotti, and
  Reining}}]{guzzo2011valence}
\bibinfo{author}{\bibfnamefont{M.}~\bibnamefont{Guzzo}},
  \bibinfo{author}{\bibfnamefont{G.}~\bibnamefont{Lani}},
  \bibinfo{author}{\bibfnamefont{F.}~\bibnamefont{Sottile}},
  \bibinfo{author}{\bibfnamefont{P.}~\bibnamefont{Romaniello}},
  \bibinfo{author}{\bibfnamefont{M.}~\bibnamefont{Gatti}},
  \bibinfo{author}{\bibfnamefont{J.~J.} \bibnamefont{Kas}},
  \bibinfo{author}{\bibfnamefont{J.~J.} \bibnamefont{Rehr}},
  \bibinfo{author}{\bibfnamefont{M.~G.} \bibnamefont{Silly}},
  \bibinfo{author}{\bibfnamefont{F.}~\bibnamefont{Sirotti}}, \bibnamefont{and}
  \bibinfo{author}{\bibfnamefont{L.}~\bibnamefont{Reining}},
  \bibinfo{journal}{Physical review letters} \textbf{\bibinfo{volume}{107}},
  \bibinfo{pages}{166401} (\bibinfo{year}{2011}).

\bibitem[{\citenamefont{Lischner et~al.}(2013)\citenamefont{Lischner,
  Vigil-Fowler, and Louie}}]{lischner2013physical}
\bibinfo{author}{\bibfnamefont{J.}~\bibnamefont{Lischner}},
  \bibinfo{author}{\bibfnamefont{D.}~\bibnamefont{Vigil-Fowler}},
  \bibnamefont{and} \bibinfo{author}{\bibfnamefont{S.~G.} \bibnamefont{Louie}},
  \bibinfo{journal}{Physical Review Letters} \textbf{\bibinfo{volume}{110}},
  \bibinfo{pages}{146801} (\bibinfo{year}{2013}).

\bibitem[{\citenamefont{Lischner et~al.}(2014)\citenamefont{Lischner,
  Vigil-Fowler, and Louie}}]{lischner2014satellite}
\bibinfo{author}{\bibfnamefont{J.}~\bibnamefont{Lischner}},
  \bibinfo{author}{\bibfnamefont{D.}~\bibnamefont{Vigil-Fowler}},
  \bibnamefont{and} \bibinfo{author}{\bibfnamefont{S.~G.} \bibnamefont{Louie}},
  \bibinfo{journal}{Physical Review B} \textbf{\bibinfo{volume}{89}},
  \bibinfo{pages}{125430} (\bibinfo{year}{2014}).

\bibitem[{\citenamefont{Hybertsen and Louie}(1986)}]{hybertsen1986electron}
\bibinfo{author}{\bibfnamefont{M.~S.} \bibnamefont{Hybertsen}}
  \bibnamefont{and} \bibinfo{author}{\bibfnamefont{S.~G.} \bibnamefont{Louie}},
  \bibinfo{journal}{Physical Review B} \textbf{\bibinfo{volume}{34}},
  \bibinfo{pages}{5390} (\bibinfo{year}{1986}).

\bibitem[{\citenamefont{Steinhoff et~al.}(2014)\citenamefont{Steinhoff,
  Rösner, Jahnke, Wehling, and Gies}}]{teinhoff2014influence}
\bibinfo{author}{\bibfnamefont{A.}~\bibnamefont{Steinhoff}},
  \bibinfo{author}{\bibfnamefont{M.}~\bibnamefont{Rösner}},
  \bibinfo{author}{\bibfnamefont{F.}~\bibnamefont{Jahnke}},
  \bibinfo{author}{\bibfnamefont{T.~O.} \bibnamefont{Wehling}},
  \bibnamefont{and} \bibinfo{author}{\bibfnamefont{C.}~\bibnamefont{Gies}},
  \bibinfo{journal}{Nano Letters} \textbf{\bibinfo{volume}{0}},
  \bibinfo{pages}{null} (\bibinfo{year}{2014}),
  \eprint{http://pubs.acs.org/doi/pdf/10.1021/nl500595u},
  \urlprefix\url{http://pubs.acs.org/doi/abs/10.1021/nl500595u}.

\bibitem[{\citenamefont{Hedin}(1965)}]{hedin1965new}
\bibinfo{author}{\bibfnamefont{L.}~\bibnamefont{Hedin}},
  \bibinfo{journal}{Physical Review} \textbf{\bibinfo{volume}{139}},
  \bibinfo{pages}{A796} (\bibinfo{year}{1965}).

\bibitem[{\citenamefont{Rozzi et~al.}(2006)\citenamefont{Rozzi, Varsano,
  Marini, Gross, and Rubio}}]{rozzi2006exact}
\bibinfo{author}{\bibfnamefont{C.~A.} \bibnamefont{Rozzi}},
  \bibinfo{author}{\bibfnamefont{D.}~\bibnamefont{Varsano}},
  \bibinfo{author}{\bibfnamefont{A.}~\bibnamefont{Marini}},
  \bibinfo{author}{\bibfnamefont{E.~K.} \bibnamefont{Gross}}, \bibnamefont{and}
  \bibinfo{author}{\bibfnamefont{A.}~\bibnamefont{Rubio}},
  \bibinfo{journal}{Physical Review B} \textbf{\bibinfo{volume}{73}},
  \bibinfo{pages}{205119} (\bibinfo{year}{2006}).

\bibitem[{\citenamefont{Ismail-Beigi}(2006)}]{ismail2006truncation}
\bibinfo{author}{\bibfnamefont{S.}~\bibnamefont{Ismail-Beigi}},
  \bibinfo{journal}{Physical Review B} \textbf{\bibinfo{volume}{73}},
  \bibinfo{pages}{233103} (\bibinfo{year}{2006}).

\bibitem[{sup(0)}]{supp}
\bibinfo{journal}{See Supplemental Material}  (\bibinfo{year}{0}).

\bibitem[{\citenamefont{Deslippe et~al.}(2012)\citenamefont{Deslippe,
  Samsonidze, Strubbe, Jain, Cohen, and Louie}}]{deslippe2012berkeleygw}
\bibinfo{author}{\bibfnamefont{J.}~\bibnamefont{Deslippe}},
  \bibinfo{author}{\bibfnamefont{G.}~\bibnamefont{Samsonidze}},
  \bibinfo{author}{\bibfnamefont{D.~A.} \bibnamefont{Strubbe}},
  \bibinfo{author}{\bibfnamefont{M.}~\bibnamefont{Jain}},
  \bibinfo{author}{\bibfnamefont{M.~L.} \bibnamefont{Cohen}}, \bibnamefont{and}
  \bibinfo{author}{\bibfnamefont{S.~G.} \bibnamefont{Louie}},
  \bibinfo{journal}{Computer Physics Communications}
  \textbf{\bibinfo{volume}{183}}, \bibinfo{pages}{1269} (\bibinfo{year}{2012}).

\bibitem[{\citenamefont{Giuliani}(2005)}]{giuliani2005quantum}
\bibinfo{author}{\bibfnamefont{G.}~\bibnamefont{Giuliani}},
  \emph{\bibinfo{title}{Quantum theory of the electron liquid}}
  (\bibinfo{publisher}{Cambridge University Press}, \bibinfo{year}{2005}).

\bibitem[{\citenamefont{Oschlies et~al.}(1995)\citenamefont{Oschlies, Godby,
  and Needs}}]{oschlies1995gw}
\bibinfo{author}{\bibfnamefont{A.}~\bibnamefont{Oschlies}},
  \bibinfo{author}{\bibfnamefont{R.}~\bibnamefont{Godby}}, \bibnamefont{and}
  \bibinfo{author}{\bibfnamefont{R.}~\bibnamefont{Needs}},
  \bibinfo{journal}{Physical Review B} \textbf{\bibinfo{volume}{51}},
  \bibinfo{pages}{1527} (\bibinfo{year}{1995}).

\bibitem[{\citenamefont{Qiu et~al.}(2013)\citenamefont{Qiu, Felipe, and
  Louie}}]{qiu2013optical}
\bibinfo{author}{\bibfnamefont{D.~Y.} \bibnamefont{Qiu}},
  \bibinfo{author}{\bibfnamefont{H.}~\bibnamefont{Felipe}}, \bibnamefont{and}
  \bibinfo{author}{\bibfnamefont{S.~G.} \bibnamefont{Louie}},
  \bibinfo{journal}{Physical review letters} \textbf{\bibinfo{volume}{111}},
  \bibinfo{pages}{216805} (\bibinfo{year}{2013}).

\bibitem[{\citenamefont{Liang et~al.}(2013)\citenamefont{Liang, Huang,
  Soklaski, and Yang}}]{liang2013quasiparticle}
\bibinfo{author}{\bibfnamefont{Y.}~\bibnamefont{Liang}},
  \bibinfo{author}{\bibfnamefont{S.}~\bibnamefont{Huang}},
  \bibinfo{author}{\bibfnamefont{R.}~\bibnamefont{Soklaski}}, \bibnamefont{and}
  \bibinfo{author}{\bibfnamefont{L.}~\bibnamefont{Yang}},
  \bibinfo{journal}{Applied Physics Letters} \textbf{\bibinfo{volume}{103}},
  \bibinfo{pages}{042106} (\bibinfo{year}{2013}).

\bibitem[{\citenamefont{Bruneval et~al.}(2006)\citenamefont{Bruneval, Vast, and
  Reining}}]{bruneval2006effect}
\bibinfo{author}{\bibfnamefont{F.}~\bibnamefont{Bruneval}},
  \bibinfo{author}{\bibfnamefont{N.}~\bibnamefont{Vast}}, \bibnamefont{and}
  \bibinfo{author}{\bibfnamefont{L.}~\bibnamefont{Reining}},
  \bibinfo{journal}{Physical Review B} \textbf{\bibinfo{volume}{74}},
  \bibinfo{pages}{045102} (\bibinfo{year}{2006}).

\bibitem[{\citenamefont{van Schilfgaarde et~al.}(2006)\citenamefont{van
  Schilfgaarde, Kotani, and Faleev}}]{van2006quasiparticle}
\bibinfo{author}{\bibfnamefont{M.}~\bibnamefont{van Schilfgaarde}},
  \bibinfo{author}{\bibfnamefont{T.}~\bibnamefont{Kotani}}, \bibnamefont{and}
  \bibinfo{author}{\bibfnamefont{S.}~\bibnamefont{Faleev}},
  \bibinfo{journal}{Physical review letters} \textbf{\bibinfo{volume}{96}},
  \bibinfo{pages}{226402} (\bibinfo{year}{2006}).

\bibitem[{\citenamefont{Hedin}(1999)}]{hedin1999correlation}
\bibinfo{author}{\bibfnamefont{L.}~\bibnamefont{Hedin}},
  \bibinfo{journal}{Journal of Physics: Condensed Matter}
  \textbf{\bibinfo{volume}{11}}, \bibinfo{pages}{R489} (\bibinfo{year}{1999}).

\bibitem[{\citenamefont{Sarma and Hwang}(2013)}]{sarma2013velocity}
\bibinfo{author}{\bibfnamefont{S.~D.} \bibnamefont{Sarma}} \bibnamefont{and}
  \bibinfo{author}{\bibfnamefont{E.}~\bibnamefont{Hwang}},
  \bibinfo{journal}{Physical Review B} \textbf{\bibinfo{volume}{87}},
  \bibinfo{pages}{045425} (\bibinfo{year}{2013}).

\bibitem[{\citenamefont{Yang et~al.}(2009)\citenamefont{Yang, Deslippe, Park,
  Cohen, and Louie}}]{yang2009excitonic}
\bibinfo{author}{\bibfnamefont{L.}~\bibnamefont{Yang}},
  \bibinfo{author}{\bibfnamefont{J.}~\bibnamefont{Deslippe}},
  \bibinfo{author}{\bibfnamefont{C.-H.} \bibnamefont{Park}},
  \bibinfo{author}{\bibfnamefont{M.~L.} \bibnamefont{Cohen}}, \bibnamefont{and}
  \bibinfo{author}{\bibfnamefont{S.~G.} \bibnamefont{Louie}},
  \bibinfo{journal}{Physical review letters} \textbf{\bibinfo{volume}{103}},
  \bibinfo{pages}{186802} (\bibinfo{year}{2009}).

\bibitem[{\citenamefont{Giannozzi et~al.}(2009)\citenamefont{Giannozzi, Baroni,
  Bonini, Calandra, Car, Cavazzoni, Ceresoli, Chiarotti, Cococcioni, Dabo
  et~al.}}]{giannozzi2009quantum}
\bibinfo{author}{\bibfnamefont{P.}~\bibnamefont{Giannozzi}},
  \bibinfo{author}{\bibfnamefont{S.}~\bibnamefont{Baroni}},
  \bibinfo{author}{\bibfnamefont{N.}~\bibnamefont{Bonini}},
  \bibinfo{author}{\bibfnamefont{M.}~\bibnamefont{Calandra}},
  \bibinfo{author}{\bibfnamefont{R.}~\bibnamefont{Car}},
  \bibinfo{author}{\bibfnamefont{C.}~\bibnamefont{Cavazzoni}},
  \bibinfo{author}{\bibfnamefont{D.}~\bibnamefont{Ceresoli}},
  \bibinfo{author}{\bibfnamefont{G.~L.} \bibnamefont{Chiarotti}},
  \bibinfo{author}{\bibfnamefont{M.}~\bibnamefont{Cococcioni}},
  \bibinfo{author}{\bibfnamefont{I.}~\bibnamefont{Dabo}}, \bibnamefont{et~al.},
  \bibinfo{journal}{Journal of Physics: Condensed Matter}
  \textbf{\bibinfo{volume}{21}}, \bibinfo{pages}{395502}
  (\bibinfo{year}{2009}).

\end{thebibliography}

\end{document}